\documentclass{article}
\usepackage{amssymb}
\usepackage{amsmath}
\usepackage{amsfonts}
\usepackage{sw20ssur}
\usepackage{cite}
\usepackage{afterpage}

\input{tcilatex}
\begin{document}

\title{{\Large On the equivalence between non-factorizable mixed-strategy
classical games and quantum games}}
\author{Azhar Iqbal, James M. Chappell, and Derek Abbott \\
{\small School of Electrical \& Electronic Engineering, the University of
Adelaide,}\\
{\small South Australia 5005, Australia.}}
\maketitle

\begin{abstract}
A game-theoretic setting provides a mathematical basis for analysis of
strategic interaction among competing agents and provides insights into both
classical and quantum decision theory and questions of strategic choice. An
outstanding mathematical question, is to understand the conditions under
which a classical game-theoretic setting can be transformed to a quantum
game, and under which conditions there is an equivalence. In this paper, we
consider quantum games as those that allow non-factorizable probabilities.
We discuss two approaches for obtaining a non-factorizable game and study
the outcome of such games. We demonstrate how the standard version of a
quantum game can be analyzed as a non-factorizable game and determine the
limitations of our approach.
\end{abstract}

\section*{Introduction}

The realization has now emerged \cite{Wilde,Nielsen,Jones} that the
processing of information cannot be separated from the underlying
fundamental physics and that the physical aspects of information processing
must be taken into consideration. This has led to a new understanding of the
information processing, cryptography, and the methods and techniques used
for communication that are based on the rules of quantum theory \cite{Peres}.

A venue where information plays a central role is in the established branch
of mathematics called game theory \cite{Binmore,Rasmusen,Osborne} that
provides the necessary mathematical tools, methods, and solution concepts
for the analysis of conflict. The understanding of games is based on some
fundamental assumptions relating to what information is available to each
participating player at a particular stage of the game. The finding that
physical aspects can have crucial role in information processing has natural
implications for player strategies and the considerations of underlying
fundamental physics become important for game theory. Such questions have
led to creation of the research area of quantum games \cite%
{MeyerDavid,EWL,EW,Boukas,Vaidman,BenjaminHayden,EnkPike,Caves,Johnson,MarinattoWeber,IqbalToor1,DuLi,Du,Piotrowski,IqbalToor3,FlitneyAbbott1,IqbalToor2,Piotrowski1,Shimamura1, FlitneyAbbott2,DuNplayer,Hanetal,IqbalWeigert,Mendes,CheonTsutsui,IqbalEPR,NawazToor,OzdemirA,Cheon,Shimamura,ChenWang,Schmidetal,IchikawaTsutsui,CheonIqbal,Ozdemir,FlitneyGreentree,Prevedel,IqbalCheon,IchikawaTsutsuiCheon,Ramzan,IqbalChapter,FlitneyHollenberg,Aharon,Bleiler,AhmedBleilerKhan,Qiang,Qiang1,ChappellA,ChappellD,IqbalAbbott,Chappell,IqbalCheonAbbott,Kolenderski,ChappellB,ChappellC,Pawela,Pawela1,Brunner,Frackiewicz,IqbalBayes,Ramanathan,Buscemi,Mohammad}%
. What these studies have shown is that quantum strategies can result in
outcomes that often defy our classical intuition.

The emergent field of quantum game theory is rapidly growing \cite%
{GoogleScholar}. Quantum game theory has two branches: (i) games based on
quantum coin tossing that explore the theory of quantum walks, and (ii)
strategic games, in the von Neumann sense, which explore quantum decision
spaces in situations of conflict. There are a number of directions that have
motivated research in quantum game theory. First and foremost, it must be
recognized that the field is a fundamental exploration of scientific
curiosity and that it provides a glimpse into quantum dynamics. Quantum
games provide mathematical settings for exploring competing interactions
e.g. between classical and quantum agents \cite{MeyerDavid,EnkPike,Aharon},
between players in the Prisoners' Dilemma game \cite%
{EWL,EW,IchikawaTsutsui,IchikawaTsutsuiCheon}, multiplayer quantum games 
\cite%
{Vaidman,DuLi,DuNplayer,Hanetal,ChenWang,Schmidetal,FlitneyGreentree,ChappellD,ChappellC}%
, interactions on quantum networks \cite{Qiang,Qiang1,Pawela}, to name a
few. Moreover, a number of authors are using a game-theoretic setting to
explore the possibility of new quantum algorithms and protocols \cite%
{MeyerDavid,Qiang,Meyer,Romanelli,LeeCF}. The area of quantum auctions \cite%
{Hogg,YangYG,WenJie} is an example of this, providing new motivation for
quantum computational and quantum network hardware. Thus far, a number of
simpler quantum games have been implemented in hardware, demonstrating
future promise \cite{Du,Schmidetal,Prevedel,IqbalEPR,Kolenderski,ChenKY}.

The speed with which new quantum technologies are emerging suggests that
soon it would be usual to take full advantage of quantum theory, and using
quantum strategies, in order to beat an opposing player at some realistic
game that uses quantum technology \cite{Aharon}. Also, there are suggestions
that quantum games can potentially provide new insights into the rise of
complexity and self-organization at the molecular level where the rules are
dictated by quantum mechanics \cite{EWL,IqbalChapter}.

In game theory, some of the simplest games to analyze are the bimatrix
games. In the area of quantum games, a well known quantization scheme for
bimatrix games was proposed by Eisert et al. \cite{EWL}. In this scheme, the
players's strategies or actions, are particular local unitary
transformations performed on an initial maximally entangled state $%
\left\vert \psi _{i}\right\rangle $ in $2\otimes 2$ Hilbert space. After the
players' actions, the quantum state passes through an unentangling gate and
thereafter is called the final state $\left\vert \psi _{f}\right\rangle $.
This state is subsequently measured using Stern-Gerlach type detectors
generating four quantum probabilities \cite{Peres}. Players' payoff
relations are expressed in terms of the payoff entries of the corresponding
bimatrix and the obtained quantum probabilities. Some experimental
realizations of quantum games are described in Refs. \cite%
{Du,Prevedel,Kolenderski}. The measurement basis for the final state $%
\left\vert \psi _{f}\right\rangle $ is defined \cite{EWL,EW} by associating
pure classical strategies of the players with the four basis vectors of the
two-qubit quantum state. Players' payoff relations contain four quantum
probabilities obtained by projecting the final quantum state onto the basis
vectors and are expressed in terms of players' local unitary transformations
and the projection postulate of quantum mechanics.

The consideration of four quantum probabilities in the players' payoff
relations leads one to ask whether the payoff relations in the quantum game
can be described as mixed-strategy payoff relations in the classical game.
This is the case when quantum probabilities are factorizable and then one
can express quantum probabilities in terms of players' mixed strategies. The
relations describing factorizability of quantum probabilities are a set of
equations that link players' mixed strategies to a probability distribution
on players' payoffs.

A set of quantum probabilities can be non-factorizable and thus cannot be
obtained from the players' mixed strategies. A quantum game can then be
described as the game in which non-factorizable probabilities are permitted.
Approaching a quantum game from this perspective, this paper presents two
approaches in obtaining games with non-factorizable probabilities. In our
first approach, factorizability is controlled by an external parameter $k\in
\lbrack 0,1]$ in the sense that assigning $k=0$ results in the factorizable
game whereas assigning $k\neq 0$ leads to a non-factorizable game. We then
ask whether a general quantum probability distribution can be described in
this way. In our second approach, we reexpess the players' payoffs relations
in a form that allows us to obtain a non-factorizble game directly from the
factorizable game by defining a function of players' strategies that
satisfies certain constraints.

\section*{Two-player quantum games}

Consider a symmetric bimatrix game \cite{Binmore,Rasmusen,Osborne}

\begin{equation}
\begin{array}{c}
\text{Alice}%
\end{array}%
\begin{array}{c}
S_{1} \\ 
S_{2}%
\end{array}%
\overset{\overset{%
\begin{array}{c}
\text{Bob}%
\end{array}%
}{%
\begin{array}{ccc}
S_{1}^{\prime } &  & S_{2}^{\prime }%
\end{array}%
}}{%
\begin{tabular}{ll}
$(\alpha ,\alpha )$ & $(\beta ,\gamma )$ \\ 
$(\gamma ,\beta )$ & $(\delta ,\delta )$%
\end{tabular}%
}  \label{Game matrix}
\end{equation}%
where $\alpha ,$ $\beta ,$ $\gamma ,$ and $\delta $ are real numbers. Assume
that Alice's and Bob's mixed strategies are $p,$ $q\in \lbrack 0,1],$
respectively, at which their payoffs can be written as

\begin{eqnarray}
\Pi _{A}(p,q) &=&\alpha pq+\beta p(1-q)+\gamma (1-p)q+\delta (1-p)(1-q), 
\notag \\
\Pi _{B}(p,q) &=&\alpha pq+\gamma p(1-q)+\beta (1-p)q+\delta (1-p)(1-q).
\label{Payoffs}
\end{eqnarray}%
A Nash equilibrium (NE)\ consists of the pair $(p^{\ast },q^{\ast })$ of
strategies such that no player has any motivation to unilaterally deviate
from it. The game's Nash inequalities take the form

\begin{equation}
\Pi _{A}(p^{\ast },q^{\ast })-\Pi _{A}(p,q^{\ast })\geq 0,\text{ \ \ }\Pi
_{B}(p^{\ast },q^{\ast })-\Pi _{B}(p^{\ast },q)\geq 0,
\end{equation}%
which take the following form for the game defined by matrix (\ref{Game
matrix})

\begin{eqnarray}
\left[ (\alpha -\beta -\gamma +\delta )q^{\ast }+(\beta -\delta )\right]
(p^{\ast }-p) &\geq &0,  \notag \\
\left[ (\alpha -\beta -\gamma +\delta )p^{\ast }+(\beta -\delta )\right]
(q^{\ast }-q) &\geq &0,  \label{NashClassGame}
\end{eqnarray}%
that gives the classical mixed strategy description of the matrix game (\ref%
{Game matrix}).

Now consider the game (\ref{Game matrix}) when played in Eisert et al.'s
quantization scheme \cite{EWL,EW}. The scheme uses two qubits to play a
quantum version of the game (\ref{Game matrix}), whose quantum state is in $%
2\otimes 2$ dimensional Hilbert space. In view of the game (\ref{Game matrix}%
), a measurement basis for quantum state of two qubits is chosen as $%
\left\vert S_{1}S_{1}^{\prime }\right\rangle ,$ $\left\vert
S_{1}S_{2}^{\prime }\right\rangle ,$ $\left\vert S_{2}S_{1}^{\prime
}\right\rangle ,$ $\left\vert S_{2}S_{2}^{\prime }\right\rangle $. An
entangled initial quantum state $\left\vert \psi _{i}\right\rangle $ is
obtained by using a two-qubit entangling gate $\hat{J}$ i.e. $\left\vert
\psi _{i}\right\rangle =\hat{J}\left\vert S_{1}S_{1}^{\prime }\right\rangle $
where $\hat{J}=\exp \left\{ i\gamma S_{2}\otimes S_{2}^{\prime }/2\right\} $
and $\gamma $ $\in \lbrack 0,\pi /2]$ is a measure of the game's
entanglement. For a separable or product game $\gamma =0$ whereas for a
maximally entangled game $\gamma =\pi /2$. The scheme considers the initial
state being a maximally entangled state $\left\vert \psi _{i}\right\rangle $%
. Players perform their local unitary transformations $\hat{U}_{A}$ and $%
\hat{U}_{B}$ from two sets of unitary transformations

\begin{eqnarray}
U(\theta ) &=&\left( 
\begin{array}{cc}
\cos (\theta /2) & \sin (\theta /2) \\ 
\text{-}\sin (\theta /2) & \cos (\theta /2)%
\end{array}%
\right)  \label{OneParameterSet} \\
U(\theta ,\phi ) &=&\left( 
\begin{tabular}{ll}
e$^{i\phi }\cos (\theta /2)$ & $\sin (\theta /2)$ \\ 
$\text{-}\sin (\theta /2)$ & e$^{-i\phi }\cos (\theta /2)$%
\end{tabular}%
\ \right) ,  \label{TwoParameterSet}
\end{eqnarray}%
where $0\leq \theta \leq \pi $ and $0\leq \phi \leq \pi /2$. These actions
change the initial maximally entangled state $\left\vert \psi
_{i}\right\rangle $ to $(\hat{U}_{A}\otimes \hat{U}_{B})\hat{J}\left\vert
S_{1}S_{1}^{\prime }\right\rangle $ which then is passed through an
untangling gate $\hat{J}^{\dagger }$ and the state of the game changes to
the final state i.e. $\left\vert \psi _{f}\right\rangle =\hat{J}^{\dagger }(%
\hat{U}_{A}\otimes \hat{U}_{B})\hat{J}\left\vert S_{1}S_{1}^{\prime
}\right\rangle $. The state $\left\vert \psi _{f}\right\rangle $ is measured
in the basis $\left\vert S_{1}S_{1}^{\prime }\right\rangle ,$ $\left\vert
S_{1}S_{2}^{\prime }\right\rangle ,$ $\left\vert S_{2}S_{1}^{\prime
}\right\rangle ,$ $\left\vert S_{2}S_{2}^{\prime }\right\rangle $ and using
the quantum probability rule, the players' payoffs are obtained as

\begin{eqnarray}
\Pi _{A}(\hat{U}_{A},\hat{U}_{B}) &=&\alpha \left\vert \left\langle
S_{1}S_{1}^{\prime }\mid \psi _{f}\right\rangle \right\vert ^{2}+\beta
\left\vert \left\langle S_{1}S_{2}^{\prime }\mid \psi _{f}\right\rangle
\right\vert ^{2}+\gamma \left\vert \left\langle S_{2}S_{1}^{\prime }\mid
\psi _{f}\right\rangle \right\vert ^{2}+\delta \left\vert \left\langle
S_{2}S_{2}^{\prime }\mid \psi _{f}\right\rangle \right\vert ^{2},  \notag \\
\Pi _{B}(\hat{U}_{A},\hat{U}_{B}) &=&\alpha \left\vert \left\langle
S_{1}S_{1}^{\prime }\mid \psi _{f}\right\rangle \right\vert ^{2}+\gamma
\left\vert \left\langle S_{1}S_{2}^{\prime }\mid \psi _{f}\right\rangle
\right\vert ^{2}+\beta \left\vert \left\langle S_{2}S_{1}^{\prime }\mid \psi
_{f}\right\rangle \right\vert ^{2}+\delta \left\vert \left\langle
S_{2}S_{2}^{\prime }\mid \psi _{f}\right\rangle \right\vert ^{2}.  \notag \\
&&  \label{EWLpayoffs}
\end{eqnarray}%
The NE for the quantum game consists of a pair $(\hat{U}_{A}^{\ast },\hat{U}%
_{B}^{\ast })$ of local unitary transformations that are obtained from the
inequalities

\begin{equation}
\Pi _{A}(\hat{U}_{A}^{\ast },\hat{U}_{B}^{\ast })-\Pi _{A}(\hat{U}_{A},\hat{U%
}_{B}^{\ast })\geq 0,\text{ \ \ }\Pi _{B}(\hat{U}_{A}^{\ast },\hat{U}%
_{B}^{\ast })-\Pi _{B}(\hat{U}_{A}^{\ast },\hat{U}_{B})\geq 0.
\end{equation}%
That is, the NE consists a pair of two-parameter unitary transformations (%
\ref{TwoParameterSet}), corresponding to the two players, such that neither
player is left with any motivation to deviate from it. For $\alpha =3,$ $%
\beta =0,$ $\gamma =5,$ $\delta =1$ the matrix (\ref{Game matrix}) gives the
game of Prisoners' Dilemma for which a Pareto-optimal NE is obtained as \cite%
{EWL}

\begin{equation}
\hat{Q}=\hat{U}_{A}(0,\pi /2)=\hat{U}_{B}(0,\pi /2)  \label{QStrategy}
\end{equation}%
at which the players' payoffs are $\Pi _{A}(\hat{Q},\hat{Q})=\Pi _{B}(\hat{Q}%
,\hat{Q})=3.$\emph{\ }Thus the quantum strategy $\hat{Q}\sim \hat{U}(0,\pi
/2)$ emerges as the new equilibrium, when both players have access to the
two-parameter set (\ref{TwoParameterSet}) of unitary $2\otimes 2$ operators.

Benjamin and Hayden \cite{BenjaminHayden} observed that when their
two-parameter set is extended to include all local unitary operations (i.e.
all of $SU(2)$) the strategy $\hat{Q}$ does not remain an equilibrium. They
showed that in the full space of deterministic quantum strategies there
exists no equilibrium for the quantum Prisoners' Dilemma. Also, they
observed that the set (\ref{TwoParameterSet}) of two-parameter quantum
strategies is not closed under composition, although this closure is the
necessary requirement for \emph{any} set of quantum strategies. It can be
explained as follows. Eisert et al. \cite{EWL,EW} permitted both players the
same strategy set but introduced an arbitrary constraint into that set. This
amounts to permitting a certain strategy while forbidding the logical
counter strategy which one would intuitively expect to be equally allowed.
Benjamin and Hayden showed \cite{BenjaminHayden} that $\hat{Q}$ emerges as
the ideal strategy only because of restricting the strategy set arbitrarily.

Not withstanding the above observations, we note that in the two-player
quantum game, pairs of unitary transformations are players' strategies that
are mapped to a set of four quantum probabilities that are normalized to 1
and in terms of which the players' payoff relations are expressed. This
mapping is achieved via the quantum probability rule \cite{Peres} that
obtains a quantum probability by squaring the modulus of the projections of
the final quantum state to a basis state in the Hilbert space. This mapping
re-expresses the four quantum probabilities in terms of the pair of players'
unitary transformations. This re-expression opens up the route to finding
Nash equilibria of the game as pairs of unitary transformations. In the
following, we revise quantization of a two-player game, discuss a
factorizable game, and present the two approaches that lead to obtaining a
non-factorizable game.

\section*{Factorizability of a set of quantum probabilities}

We notice that the payoff relations (\ref{EWLpayoffs}) can be written as

\begin{eqnarray}
\Pi _{A}(\hat{U}_{A},\hat{U}_{B}) &=&\alpha \epsilon _{1}+\beta \epsilon
_{2}+\gamma \epsilon _{3}+\delta \epsilon _{4},  \notag \\
\Pi _{B}(\hat{U}_{A},\hat{U}_{B}) &=&\alpha \epsilon _{1}+\gamma \epsilon
_{2}+\beta \epsilon _{3}+\delta \epsilon _{4},  \label{QPayoffs}
\end{eqnarray}%
where

\begin{eqnarray}
\epsilon _{1} &=&\left\vert \left\langle S_{1}S_{1}^{\prime }\mid \psi
_{f}\right\rangle \right\vert ^{2},\text{ }\epsilon _{2}=\left\vert
\left\langle S_{1}S_{2}^{\prime }\mid \psi _{f}\right\rangle \right\vert
^{2},  \notag \\
\epsilon _{3} &=&\left\vert \left\langle S_{2}S_{1}^{\prime }\mid \psi
_{f}\right\rangle \right\vert ^{2},\text{ }\epsilon _{4}=\left\vert
\left\langle S_{2}S_{2}^{\prime }\mid \psi _{f}\right\rangle \right\vert
^{2},  \label{upsilons}
\end{eqnarray}%
are four quantum probabilities obtained using the quantum probability rule
i.e.~projecting the final quantum state $\left\vert \psi _{f}\right\rangle $
of the game to the four basis vectors $\left\vert S_{1}S_{1}^{\prime
}\right\rangle ,$ $\left\vert S_{1}S_{2}^{\prime }\right\rangle ,$ $%
\left\vert S_{2}S_{1}^{\prime }\right\rangle ,$ $\left\vert
S_{2}S_{2}^{\prime }\right\rangle $. The probabilities $\epsilon _{i}$ are
normalized i.e.

\begin{equation}
\Sigma _{i=1}^{4}\epsilon _{i}=1.  \label{Normalization}
\end{equation}

Comparing the payoff relations in the classical game (\ref{Payoffs}) and in
the quantum game (\ref{QPayoffs}), we notice that the payoff relations in
the quantum game can be reduced to the payoff relations in the classical
game when probabilities $\epsilon _{i}$ are factorizable. Probabilities $%
\epsilon _{i}$ are factorizable when for a given set of values assigned in
range $[0,1]$ to probabilities $\epsilon _{i},$ we can find two
probabilities $p,q\in \lbrack 0,1]$ such that $\epsilon _{i}$ can be written
in terms of $p$ and $q$ as

\begin{equation}
\epsilon _{1}=pq,\text{ }\epsilon _{2}=p(1-q),\text{ }\epsilon _{3}=(1-p)q,%
\text{ }\epsilon _{4}=(1-p)(1-q).  \label{factorizability}
\end{equation}%
That is, in the classical mixed-strategy version of the two-player game, the
four probabilities $\epsilon _{i}$ appearing in the payoff relations (\ref%
{Payoffs}) are re-expressed in terms of players' strategic variables $p$ and 
$q$ via the factorizability conditions (\ref{factorizability}). If this is
the case for the probabilities $\epsilon _{i}$ then we can associate the
probabilities $p$ and $q$ to the players Alice and Bob, respectively, so
that the payoff relations in the quantum game (\ref{QPayoffs}) are
interpreted as corresponding to a mixed-strategy classical game i.e.

\begin{equation}
\Pi _{A}(\hat{U}_{A},\hat{U}_{B})=\Pi _{A}(p,q),\text{ \ \ }\Pi _{B}(\hat{U}%
_{A},\hat{U}_{B})=\Pi _{B}(p,q),  \label{QPayoffs1}
\end{equation}%
with $p$ and $q$ satisfying the factorizability relations (\ref%
{factorizability}). In (\ref{factorizability}) $\epsilon _{i}$ represent not
just a specific set of four numbers in $[0,1]$ that add up to $1$, but the
entirety of such four numbers that can be generated by the quantum
mechanical setup used for playing a quantum game. As it can be seen from the
Eq.~(\ref{QPayoffs}), in Eisert et al's scheme, the full range of the
players' payoffs become accessible by giving players access to unitary
transformations. In other words, a pair of unitary transformations results
in a normalized set of four probabilities and Nash inequalities lead us to
obtaining a pair of unitary transformations as a NE.

\section*{Games with non-factorizable probabilities}

Quantum probabilities $\epsilon _{i}$, however, may not be factorizable in
the sense described by Eq.~(\ref{factorizability}). That is, the measurement
stage of a quantum game can result in such probabilities $\epsilon _{i}$ $%
(0\leq \epsilon _{i}\leq 1)$ that one cannot find $p,q\in \lbrack 0,1]$ so
that Eqs.\textit{~}(\ref{factorizability}) are satisfied. Viewing the payoff
relations (\ref{QPayoffs}) from this probabilistic viewpoint, it then seems
natural to ask whether we can obtain the payoffs (\ref{QPayoffs}) by simply
removing the factorizability relations (\ref{factorizability}) and if this
is the case then what are the possibly new outcomes of the game. A quantum
game allows obtaining sets of non-factorizable probabilities and in this
paper we look at the role of non-factorizable probabilities on the outcome
of a game. This can also be stated as follows. Considering Eqs.~(\ref%
{QPayoffs},\ref{QPayoffs1}), we can describe the factorizable game as the
one for which

\begin{equation}
\Pi _{A}(p,q)=\alpha \epsilon _{1}+\beta \epsilon _{2}+\gamma \epsilon
_{3}+\delta \epsilon _{4},\text{ \ \ }\Pi _{B}(p,q)=\alpha \epsilon
_{1}+\gamma \epsilon _{2}+\beta \epsilon _{3}+\delta \epsilon _{4},
\label{Non-factor-payoffs}
\end{equation}%
where $\Sigma _{i=1}^{4}\epsilon _{i}=1$ and the probabilities $\epsilon
_{i} $ are related to players's strategies $p$ and $q$ via the
factorizability relations (\ref{factorizability}). In the following, we
consider two approaches in obtaining non-factorizable probabilities.

\subsection*{The first approach}

Our first approach considers an external parameter $k$ that determines
whether the probabilities $\epsilon _{i}$ $(0\leq i\leq 4)$ are factorizable
or not. For this, we consider the following probability distribution

\begin{eqnarray}
\epsilon _{1} &=&(2k-1)^{2}pq,\text{ \ }\epsilon _{2}=(1-k)p(1-q)+kq(1-p), 
\notag \\
\epsilon _{3} &=&(1-k)q(1-p)+kp(1-q),\text{ \ }\epsilon
_{4}=4k(1-k)pq+(1-p)(1-q).  \label{FirstApproach}
\end{eqnarray}%
It can be confirmed that for $k$ in the range $\left[ 0,1\right] $ we have $%
0\leq \epsilon _{i}\leq 1$ and that probabilities $\epsilon _{i}$ are
normalized according to the Eq.~(\ref{Normalization}). When $k=0$ the
distribution (\ref{FirstApproach}) reduces to the factorizable distribution
of the Eq.~(\ref{factorizability}). However, when $k$ is non-zero, the
probability distribution (\ref{FirstApproach}) is not factorizable. Using
the payoff relations (\ref{Non-factor-payoffs}), we now obtain the payoffs
for Alice and Bob for the distribution (\ref{FirstApproach}) as 
\begin{eqnarray}
\Pi _{A}(p,q,k) &=&\alpha (2k-1)^{2}pq+\beta \left[ (1-k)p(1-q)+kq(1-p)%
\right]  \notag \\
&&+\gamma \left[ (1-k)q(1-p)+kp(1-q)\right] +\delta \left[
4k(1-k)pq+(1-p)(1-q)\right] ,  \notag \\
\Pi _{B}(p,q,k) &=&\alpha (2k-1)^{2}pq+\gamma \left[ (1-k)p(1-q)+kq(1-p)%
\right]  \notag \\
&&+\beta \left[ (1-k)q(1-p)+kp(1-q)\right] +\delta \left[
4k(1-k)pq+(1-p)(1-q)\right] .  \notag \\
&&
\end{eqnarray}%
The NE strategy pair $(p^{\ast },q^{\ast })$ is then obtained from the
inequalities 
\begin{eqnarray}
\left\{ \left[ \alpha (1-2k)^{2}-\beta -\gamma +\delta \left\{
1+4k(1-k)\right\} \right] q^{\ast }+(\beta -\delta )-k(\beta -\gamma
)\right\} (p^{\ast }-p) &\geq &0,  \notag \\
\left\{ \left[ \alpha (1-2k)^{2}-\beta -\gamma +\delta \left\{
1+4k(1-k)\right\} \right] p^{\ast }+(\beta -\delta )-k(\beta -\gamma
)\right\} (q^{\ast }-q) &\geq &0.  \notag \\
&&
\end{eqnarray}%
Now, for the Prisoners' Dilemma game considered above, we have $\alpha =3,$ $%
\beta =0,$ $\gamma =5,$ $\delta =1$ and these reduce the above NE conditions
to 
\begin{eqnarray}
\left\{ -1-q+k\left[ 5-8q(1-k)\right] \right\} (p^{\ast }-p) &\geq &0, 
\notag \\
\left\{ -1-p+k\left[ 5-8p(1-k)\right] \right\} (q^{\ast }-q) &\geq &0.
\end{eqnarray}%
For $k=0$ these relations give the NE in the classical factorizable game
i.e. $(p^{\ast },q^{\ast })=(0,0)$ at which the players payoffs are obtained
as $\Pi _{A,B}(p^{\ast },q^{\ast },k)=\Pi _{A,B}(0,0,0)=1$.

Now consider the case when $k=1$ for which we find the NE being $(p^{\ast
},q^{\ast })=(1,1)$ and the players' payoffs are obtained as $\Pi
_{A,B}(p,q,k)=\Pi _{A,B}(1,1,1)=3$. These payoffs are same as obtained in
the maximally entangled quantum game in Eisert et al's scheme \cite{EWL}
with both players playing the quantum strategy $\hat{Q}$ defined in Eq.~(\ref%
{QStrategy}).

However, we notice that the probability distribution in Eq.~(\ref%
{FirstApproach}) is not as general as a quantum probability distribution can
be within a quantum game. This is because for (\ref{FirstApproach}) to be a
quantum probability distribution it is required to obey only the
normalization constraint (\ref{Normalization}). Although the probability
distribution (\ref{FirstApproach}) is normalized it also obeys other
restrictions because of its particular form and the way it is defined. This
can also be stated as follows. Whereas the probability distribution (\ref%
{FirstApproach}) is normalized, not every normalized quantum probability
distribution will have this form. That is, there can be quantum probability
distributions that cannot be written in the same form as the distribution (%
\ref{FirstApproach}). However, in spite of these limitations, the
probability distribution (\ref{FirstApproach}) demonstrates the effect of a
non-factorizable probability distribution on the outcome of a game. It also
produces the classical game as a special case. In the following, we present
a second approach that defines a probability distribution using a function
of players' strategies $p$ and $q$ that is subject to certain constraints.

\subsection*{The second approach}

At this stage, we notice that when $\epsilon _{i}$ are factorizable in the
sense described by Eqs.\textit{~}(\ref{factorizability}) that give the
relationship between $p,$ $q$ and $\epsilon _{i}$. Using these relations,
the players' strategies $p$ and $q$ can be expressed in terms of
probabilities $\epsilon _{i}$ as

\begin{equation}
p=\epsilon _{1}+\epsilon _{2},\text{ \ \ }q=\epsilon _{1}+\epsilon _{3},
\label{p&q}
\end{equation}%
and using Eqs.\textit{~}(\ref{upsilons}), we can write Eqs.~(\ref{p&q}) as

\begin{eqnarray}
p &=&\left\vert \left\langle S_{1}S_{1}^{\prime }\mid \psi _{f}\right\rangle
\right\vert ^{2}+\left\vert \left\langle S_{1}S_{2}^{\prime }\mid \psi
_{f}\right\rangle \right\vert ^{2},  \notag \\
q &=&\left\vert \left\langle S_{1}S_{1}^{\prime }\mid \psi _{f}\right\rangle
\right\vert ^{2}+\left\vert \left\langle S_{2}S_{1}^{\prime }\mid \psi
_{f}\right\rangle \right\vert ^{2}.
\end{eqnarray}%
Knowing that $p$ and $q$ are the players' strategies, and each player has
the freedom to play whatever s/he likes, the strategies $p$ and $q$ are
considered to be independent of each other.

When quantum probabilities $\epsilon _{i}$ are factorizable in the sense
described by Eqs.~(\ref{factorizability}), the payoff relations (\ref%
{QPayoffs}) can also be interpreted in terms of playing a game that involves
tossing a pair of coins as follows. Consider a pair of biased coins that are
tossed together. Either coin can land in the head $(H)$ or the tail $(T)$
state and for the pair we can define

\begin{equation}
\epsilon _{1}=\Pr (H,H),\text{ }\epsilon _{2}=\Pr (H,T),\text{ }\epsilon
_{3}=\Pr (T,H),\text{ }\epsilon _{4}=\Pr (T,T),
\end{equation}%
as being the probabilities of the coins landing in the $(H,H)$, $(H,T)$, $%
(T,H)$, $(T,T)$ states, respectively. Here, for instance, $\epsilon _{2}=\Pr
(H,T)$ is the probability that the first coin (or Alice's coin) lands in the 
$H$ state whereas the second coin (or Bob's coin) lands in the $T$ state.
Now, referring to the Eq.~(\ref{p&q}), $p$ can then be interpreted as being
the probability, in the joint toss of two coins, that the first coin lands
in the $H$ state and, likewise, $q$ can be interpreted as being the
probability that the second coin lands in the $H$ state.

However, we are interpreting $p$ and $q$ as being the players' strategies
which means that a player plays his/her strategy by changing the bias of the
coin to which he/she is given access to. Also, we notice that, for
factorizable quantum probabilities, using Eqs.~(\ref{factorizability},\ref%
{p&q}) we can write

\begin{eqnarray}
\epsilon _{1} &=&(\epsilon _{1}+\epsilon _{2})(\epsilon _{1}+\epsilon _{3}),%
\text{ }\epsilon _{2}=(\epsilon _{1}+\epsilon _{2})(1-\epsilon _{1}-\epsilon
_{3}),\text{ }  \notag \\
\epsilon _{3} &=&(1-\epsilon _{1}-\epsilon _{2})(\epsilon _{1}+\epsilon
_{3}),\text{ }\epsilon _{4}=(1-\epsilon _{1}-\epsilon _{2})(1-\epsilon
_{1}-\epsilon _{3}).  \label{factorizability2}
\end{eqnarray}

We now suggest another approach in considering the players' payoffs (\ref%
{QPayoffs}) in the quantum game. In view of the relations $p=\epsilon
_{1}+\epsilon _{2}$, $q=\epsilon _{1}+\epsilon _{3},$ and the normalization
constraint $\Sigma _{i=1}^{4}\epsilon _{i}=1$ we can rewrite the payoffs (%
\ref{QPayoffs}) as

\begin{eqnarray}
\Pi _{A}(\hat{U}_{A},\hat{U}_{B}) &=&\alpha \epsilon _{1}+\beta \epsilon
_{2}+\gamma \epsilon _{3}+\delta \epsilon _{4}  \notag \\
&=&\epsilon _{1}(\alpha -\beta -\gamma +\delta )+(\beta -\delta )(\epsilon
_{1}+\epsilon _{2})+(\gamma -\delta )(\epsilon _{1}+\epsilon _{3})+\delta 
\notag \\
&=&\epsilon _{1}(\alpha -\beta -\gamma +\delta )+(\beta -\delta )p+(\gamma
-\delta )q+\delta ,  \notag \\
\Pi _{B}(\hat{U}_{A},\hat{U}_{B}) &=&\alpha \epsilon _{1}+\gamma \epsilon
_{2}+\beta \epsilon _{3}+\delta \epsilon _{4}  \notag \\
&=&\epsilon _{1}(\alpha -\beta -\gamma +\delta )+(\gamma -\delta )p+(\beta
-\delta )q+\delta .  \label{Re-def-eqs}
\end{eqnarray}%
We also note that the players' payoffs (\ref{Payoffs}) in the factorizable
game can be expressed as

\begin{eqnarray}
\Pi _{A}(p,q) &=&\alpha pq+\beta p(1-q)+\gamma (1-p)q+\delta (1-p)(1-q) 
\notag \\
&=&pq(\alpha -\beta -\gamma +\delta )+(\beta -\delta )p+(\gamma -\delta
)q+\delta ,  \notag \\
\Pi _{B}(p,q) &=&\alpha pq+\gamma p(1-q)+\beta (1-p)q+\delta (1-p)(1-q) 
\notag \\
&=&pq(\alpha -\beta -\gamma +\delta )+(\gamma -\delta )p+(\beta -\delta
)q+\delta .  \label{Re-def-Payoffs}
\end{eqnarray}

Comparing Eqs.~(\ref{Re-def-Payoffs},\ref{Re-def-eqs}) we note that taking $%
\epsilon _{1}=pq$ makes the payoffs (\ref{Re-def-eqs}) equivalent to the
payoffs (\ref{Re-def-Payoffs}) in the factorizable game. That is, with
players' strategic variables being $p=\epsilon _{1}+\epsilon _{2}$ and $%
q=\epsilon _{1}+\epsilon _{3}$ and $\epsilon _{1}=pq,$ the players' payoffs
in the quantum game are reduced to their payoffs in the factorizable game.
Also, the product $\epsilon _{1}=pq$ takes the pair $(p,q)$ to the interval $%
\left[ 0,1\right] $ and is responsible for reducing the right sides of Eqs.~(%
\ref{Re-def-eqs}) to the right sides of Eqs.~(\ref{Re-def-Payoffs}).

This motivates considering $\epsilon _{1}$ as a function of the players'
independent strategic variables $p$ and $q$ i.e. $\epsilon _{1}=\epsilon
_{1}(p,q)$. It is a function of the players' strategic variables $p$ and $q$
that maps the pair $(p,q)$ to the interval $\left[ 0,1\right] $. Taking $%
\epsilon _{1}=\epsilon _{1}(p,q)$ allows us to consider forms of the
function $\epsilon _{1}(p,q)$ that are different from the product $pq$. In
turn, this permits departing from the factorizability relations (\ref%
{factorizability}) and thus from the factorizable game.

We also note from Eqs.~(\ref{Re-def-eqs}) that the function $\epsilon _{1}$
connects the two players' payoffs together and reminds us of the term $%
\gamma $ representing the measure of entanglement in the players' payoff
relations within the quantum game that is played in Eisert et al.'s scheme.
We write the payoff relations (\ref{Re-def-eqs}) for a non-factorizable game
as

\begin{eqnarray}
\Pi _{A}(p,q) &=&\epsilon _{1}(p,q)(\alpha -\beta -\gamma +\delta )+(\beta
-\delta )p+(\gamma -\delta )q+\delta ,  \notag \\
\Pi _{B}(p,q) &=&\epsilon _{1}(p,q)(\alpha -\beta -\gamma +\delta )+(\gamma
-\delta )p+(\beta -\delta )q+\delta ,  \label{NewGame}
\end{eqnarray}%
and consider it being the defining payoff relations for the non-factorizable
game. Note that for the quantum game the players' strategies are unitary
transformations $\hat{U}_{A}$ and $\hat{U}_{B}$, whereas for the new game
defined by payoff relations (\ref{NewGame}) the players' strategies are $%
p,q\in \lbrack 0,1]$. Also, the two players' payoffs are linked together via
the function $\epsilon _{1}=\epsilon _{1}(p,q)$. The game defined by the
payoff relations (\ref{NewGame}) makes no reference to the quantum
operations and can thus simply be called a non-factorizable game or a game
that permits non-factorizable probabilities.

With Eqs.~(\ref{NewGame}) being the defining payoff relations for our
non-factorizable game, an immediate question would be what are the
restrictions on the type of functions $\epsilon _{1}(p,q)$. To determine
this, we note that the permissible ranges of the players' payoffs in Eqs.~(%
\ref{NewGame}) and Eqs.~(\ref{QPayoffs}) must be identical. In view of the
right sides of Eqs.~(\ref{QPayoffs}), we put Eqs.~(\ref{NewGame}) in the
following form

\begin{eqnarray}
\Pi _{A}(p,q) &=&\alpha \epsilon _{1}(p,q)+\beta \left[ p-\epsilon _{1}(p,q)%
\right] +\gamma \left[ q-\epsilon _{1}(p,q)\right] +\delta \left[ \epsilon
_{1}(p,q)-(p+q)+1\right] ,  \notag \\
\Pi _{B}(p,q) &=&\alpha \epsilon _{1}(p,q)+\gamma \left[ p-\epsilon _{1}(p,q)%
\right] +\beta \left[ q-\epsilon _{1}(p,q)\right] +\delta \left[ \epsilon
_{1}(p,q)-(p+q)+1\right] ,  \notag \\
&&  \label{NewGame1}
\end{eqnarray}%
and for the right sides of these equations to be identical to the right
sides of Eqs.~(\ref{QPayoffs}) we require

\begin{equation}
\epsilon _{1}(p,q)\leq p,\text{ \ \ }\epsilon _{1}(p,q)\leq q,\text{ \ \ }%
\epsilon _{1}(p,q)\leq p+q.  \label{Restrictions}
\end{equation}%
Note that the right sides of the payoff relations (\ref{NewGame1}) are
equivalent to the right sides of the payoff relations (\ref{QPayoffs}) in
the quantum game. This can be confirmed as follows. Referring to Eqs.~(\ref%
{NewGame}), we notice that for given $p,q\in \lbrack 0,1]$ as players'
independent strategies, the function $\epsilon _{1}(p,q)$ gives a value in $%
[0,1]$, the functions $\epsilon _{2}(p,q)$ and $\epsilon _{3}(p,q)$ can then
be defined as

\begin{eqnarray}
\epsilon _{2}(p,q) &=&p-\epsilon _{1}(p,q),\text{ \ \ }\epsilon
_{3}(p,q)=q-\epsilon _{1}(p,q),  \notag \\
\epsilon _{4}(p,q) &=&1-[(p+q)-\epsilon _{1}(p,q)].  \label{UpsilonsDefined}
\end{eqnarray}%
When the restrictions (\ref{Restrictions}) hold, the functions $\epsilon
_{2}(p,q),$ $\epsilon _{3}(p,q),$ $\epsilon _{4}(p,q)$ produce values within
the range $\left[ 0,1\right] $. Of course, the function $\epsilon
_{1}(p,q)=pq$, which results in the factorizable game, satisfies these
requirements. In view of the restrictions (\ref{Restrictions}), and a
non-factorizable game for which $\epsilon _{1}(p,q)\neq pq$, we require the
function $\epsilon _{1}(p,q)$ to be restricted by the following condition

\begin{equation}
\epsilon _{1}(p,q)\leq pq.
\end{equation}%
Examples of the functions that satisfy this requirement include $\epsilon
_{1}(p,q)=(pq)^{2}$ and $\epsilon _{1}(p,q)=p^{2}q^{3}$, amongst others.

Note that, in view of (\ref{Restrictions}), the range of function $\epsilon
_{1}(p,q)$\ and the values assigned to $p$\ and $q$, as being players'
strategies, are in the interval $[0,1]$. In this case, Eqs.~(\ref%
{UpsilonsDefined}) generate values for $\epsilon _{2}(p,q),$\ $\epsilon
_{3}(p,q),$\ $\epsilon _{4}(p,q)$ in the range $[0,1]$. Also, Eqs.~(\ref%
{UpsilonsDefined}) show that $\sum\limits_{i=1}^{4}\epsilon _{i}(p,q)=1$,
and thus $\epsilon _{i}(p,q)$\ are probabilities for $1\leq i\leq 4$. The
converse is also true. That is, for given values for four normalized
probabilities $\epsilon _{i}$\ $(1\leq i\leq 4)$ using Eq.~(\ref{p&q}) we
can determine values for $p$\ and $q$\ as being $p=\epsilon _{1}+\epsilon
_{2}$\ and $q=\epsilon _{1}+\epsilon _{3}.$

\subsubsection*{Nash equilibria for the game with non-factorizable
probabilities}

The pair of strategies $(p^{\ast },q^{\ast })$ defining Nash equilibria are
obtained from the inequalities

\begin{equation}
\Pi _{A}(p^{\ast },q^{\ast })-\Pi _{A}(p,q^{\ast })\geq 0,\text{ }\Pi
_{B}(p^{\ast },q^{\ast })-\Pi _{B}(p^{\ast },q)\geq 0,
\end{equation}%
which, for the game (\ref{NewGame}), can be written as

\begin{eqnarray}
\Pi _{A}(p^{\ast },q^{\ast })-\Pi _{A}(p,q^{\ast }) &=&\left[ \epsilon
_{1}(p^{\ast },q^{\ast })-\epsilon _{1}(p,q^{\ast })\right] (\alpha -\beta
-\gamma +\delta )+(p^{\ast }-p)(\beta -\delta )\geq 0,  \notag \\
\Pi _{B}(p^{\ast },q^{\ast })-\Pi _{B}(p^{\ast },q) &=&\left[ \epsilon
_{1}(p^{\ast },q^{\ast })-\epsilon _{1}(p^{\ast },q)\right] (\alpha -\beta
-\gamma +\delta )+(q^{\ast }-q)(\beta -\delta )\geq 0.  \notag \\
&&  \label{NE-Gen-Fun}
\end{eqnarray}%
For $\epsilon _{1}(p,q)=pq$ these equations are reduced to Eqs.~(\ref%
{NashClassGame}). For $\epsilon _{1}(p,q)=p^{2}q^{2}$ the inequalities (\ref%
{NE-Gen-Fun}) give

\begin{eqnarray}
\Pi _{A}(p^{\ast },q^{\ast })-\Pi _{A}(p,q^{\ast }) &=&(p^{\ast }-p)\left[
q^{\ast 2}(p^{\ast }+p)(\alpha -\beta -\gamma +\delta )+(\beta -\delta )%
\right] \geq 0,  \notag \\
\Pi _{B}(p^{\ast },q^{\ast })-\Pi _{B}(p^{\ast },q) &=&(q^{\ast }-q)\left[
p^{\ast 2}(q^{\ast }+q)(\alpha -\beta -\gamma +\delta )+(\beta -\delta )%
\right] \geq 0,  \notag \\
&&
\end{eqnarray}%
and, likewise, for $\epsilon _{1}(p,q)=p^{2}q^{3}$ the inequalities (\ref%
{NE-Gen-Fun}) give

\begin{eqnarray}
\Pi _{A}(p^{\ast },q^{\ast })-\Pi _{A}(p,q^{\ast }) &=&(p^{\ast }-p)\left[
q^{\ast 3}(p^{\ast }+p)(\alpha -\beta -\gamma +\delta )+(\beta -\delta )%
\right] \geq 0,  \notag \\
\Pi _{B}(p^{\ast },q^{\ast })-\Pi _{B}(p^{\ast },q) &=&(q^{\ast }-q)\left[
p^{\ast 2}(q^{\ast 2}+qq^{\ast }+q^{2})(\alpha -\beta -\gamma +\delta
)+(\beta -\delta )\right] \geq 0,  \notag \\
&&
\end{eqnarray}%
Now consider the situation when the strategy pair $(p^{\ast },q^{\ast
})=(1,1)$ exists as a NE. As it is apparent from Eqs.~(\ref{NashClassGame})
that for $\epsilon _{1}(p,q)=pq$ the pair $(p^{\ast },q^{\ast })=(1,1)$
exists as NE when $(\alpha -\gamma )\geq 0.$ For $\epsilon
_{1}(p,q)=p^{2}q^{2}$ and the pair $(p^{\ast },q^{\ast })=(1,1)$ we obtain

\begin{eqnarray}
\Pi _{A}(1,1)-\Pi _{A}(p,1) &=&(1-p)\left[ (\alpha -\gamma )(1+p)+p(-\beta
+\delta )\right] \geq 0,  \notag \\
\Pi _{B}(1,1)-\Pi _{B}(1,q) &=&(1-q)\left[ (\alpha -\gamma )(1+q)+q(-\beta
+\delta )\right] \geq 0,
\end{eqnarray}%
and thus $(p^{\ast },q^{\ast })=(1,1)$ exists as a NE when, in addition to $%
(\alpha -\gamma )\geq 0,$ we also have $(\delta -\beta )\geq 0$. For $%
\epsilon _{1}(p,q)=p^{2}q^{3}$ and the same pair $(p^{\ast },q^{\ast
})=(1,1) $ we obtain

\begin{eqnarray}
\Pi _{A}(1,1)-\Pi _{A}(p,1) &=&(1-p)\left[ (\alpha -\gamma )+p(\alpha -\beta
-\gamma +\delta )\right] \geq 0,  \notag \\
\Pi _{B}(1,1)-\Pi _{B}(1,q) &=&(1-q)\left[ (\alpha -\gamma
)+(q+q^{2})(\alpha -\beta -\gamma +\delta )\right] \geq 0.
\end{eqnarray}%
That is, when both $(\alpha -\gamma )\geq 0$ and $(\delta -\beta )\geq 0$
are true, we have the strategy pair $(p^{\ast },q^{\ast })=(1,1)$ existing
as a NE for the both cases i.e. when $\epsilon _{1}(p,q)=p^{2}q^{2}$ and $%
p^{2}q^{3}$.

In this approach in extending a game from its classical mixed-strategy
version to a non-classical version, the players' strategies involve one
parameter for each i.e. $p$ and $q\in \lbrack 0,1]$. We now refer to Eisert
et al.'s quantum version of the same game \cite{EWL,EW} in which players'
payoff relations in the quantized version of the game involve one- and
two-parameter strategy sets (\ref{OneParameterSet},\ref{TwoParameterSet}).
As in our non-classical extension of a classical game the players have
one-parameter strategy sets, it is appropriate to compare our non-classical
extension above to the quantum version of Eisert et al. \cite{EWL,EW} when
it involves one-parameter strategy sets. To achieve this, and with reference
to the game (\ref{Game matrix}), we recast quantum game of Eisert et al.
with one-parameter strategy set in form of the non-classical game above.
From Ref. \cite{EW} we note that the players' payoffs for one-parameter
strategy set are

\begin{eqnarray}
\Pi _{A}(\theta _{A},\theta _{B}) &=&\alpha \left\vert \cos (\theta
_{A}/2)\cos (\theta _{B}/2)\right\vert ^{2}+\beta \left\vert \sin (\theta
_{B}/2)\cos (\theta _{A}/2)\right\vert ^{2}+  \notag \\
&&\gamma \left\vert \cos (\theta _{B}/2)\sin (\theta _{A}/2)\right\vert
^{2}+\delta \left\vert \sin (\theta _{A}/2)\sin (\theta _{B}/2)\right\vert
^{2},  \label{AlicePayoff} \\
\Pi _{A}(\theta _{A},\theta _{B}) &=&\alpha \left\vert \cos (\theta
_{A}/2)\cos (\theta _{B}/2)\right\vert ^{2}+\gamma \left\vert \sin (\theta
_{B}/2)\cos (\theta _{A}/2)\right\vert ^{2}+  \notag \\
&&\beta \left\vert \cos (\theta _{B}/2)\sin (\theta _{A}/2)\right\vert
^{2}+\delta \left\vert \sin (\theta _{A}/2)\sin (\theta _{B}/2)\right\vert
^{2}.  \label{BobPayoff}
\end{eqnarray}%
Comparing Eqs.~(\ref{NewGame1}) with Eqs.~(\ref{AlicePayoff},\ref{BobPayoff}%
) and noticing that $0\leq \theta \leq \pi $ whereas $p,$ $q\in \left[ 0,1%
\right] $, we define $p=\theta _{A}/\pi $, $q=\theta _{B}/\pi $ and

\begin{eqnarray}
\epsilon _{1}(p,q) &=&\left\vert \cos (\theta _{A}/2)\cos (\theta
_{B}/2)\right\vert ^{2},\text{ }p-\epsilon _{1}(p,q)=\left\vert \sin (\theta
_{B}/2)\cos (\theta _{A}/2)\right\vert ^{2},  \notag \\
q-\epsilon _{1}(p,q) &=&\left\vert \cos (\theta _{B}/2)\sin (\theta
_{A}/2)\right\vert ^{2},\text{ }\epsilon _{1}(p,q)-(p+q)+1=\left\vert \sin
(\theta _{A}/2)\sin (\theta _{B}/2)\right\vert ^{2},  \notag \\
&&
\end{eqnarray}%
from which one obtains

\begin{equation}
p=\cos ^{2}(\theta _{A}/2),\text{ }q=\cos ^{2}(\theta _{B}/2),\text{ }%
\epsilon _{1}(p,q)=pq,
\end{equation}%
and that gives us the factorizable game as discussed just before the Eq.~(%
\ref{NewGame}). This shows that Eisert et al.'s quantum game with
one-parameter strategy set, is identical to the classical factorizable game.
For the same quantum game with two-parameter strategy set (\ref%
{TwoParameterSet}), the players' payoffs are obtained as

\begin{eqnarray}
\Pi _{A}(\theta _{A},\phi _{A};\theta _{B},\phi _{B}) &=&\alpha \left\vert
\cos (\phi _{A}+\phi _{B})\cos (\theta _{A}/2)\cos (\theta
_{B}/2)\right\vert ^{2}+  \notag \\
&&\beta \left\vert \cos (\phi _{A})\cos (\theta _{A}/2)\sin (\theta
_{B}/2)-\sin (\phi _{B})\sin (\theta _{A}/2)\cos (\theta _{B}/2)\right\vert
^{2}+  \notag \\
&&\gamma \left\vert \sin (\phi _{A})\cos (\theta _{A}/2)\sin (\theta
_{B}/2)-\cos (\phi _{B})\sin (\theta _{A}/2)\cos (\theta _{B}/2)\right\vert
^{2}+  \notag \\
&&\delta \left\vert \sin (\phi _{A}+\phi _{B})\cos (\theta _{A}/2)\cos
(\theta _{B}/2)+\sin (\theta _{A}/2)\sin (\theta _{B}/2)\right\vert ^{2}.
\label{AlicePayoff2}
\end{eqnarray}%
As was the case for one-parameter strategy set, we now compare Eqs.~(\ref%
{NewGame1}) with Eqs.~(\ref{AlicePayoff2},\ref{BobPayoff}). Recall that $%
0\leq \theta \leq \pi $ and $0\leq \phi \leq \pi /2$ whereas $p,$ $q\in %
\left[ 0,1\right] $ we obtain

\begin{eqnarray}
\epsilon _{1}(p,q) &=&\left\vert \cos (\phi _{A}+\phi _{B})\cos (\theta
_{A}/2)\cos (\theta _{B}/2)\right\vert ^{2},  \notag \\
p-\epsilon _{1}(p,q) &=&\left\vert \cos (\phi _{A})\cos (\theta _{A}/2)\sin
(\theta _{B}/2)-\sin (\phi _{B})\sin (\theta _{A}/2)\cos (\theta
_{B}/2)\right\vert ^{2},  \notag \\
q-\epsilon _{1}(p,q) &=&\left\vert \sin (\phi _{A})\cos (\theta _{A}/2)\sin
(\theta _{B}/2)-\cos (\phi _{B})\sin (\theta _{A}/2)\cos (\theta
_{B}/2)\right\vert ^{2},  \notag \\
\epsilon _{1}(p,q)-(p+q)+1 &=&\left\vert \sin (\phi _{A}+\phi _{B})\cos
(\theta _{A}/2)\cos (\theta _{B}/2)+\sin (\theta _{A}/2)\sin (\theta
_{B}/2)\right\vert ^{2},
\end{eqnarray}%
and using these equations, $p$ and $q$ can then be expressed in terms of $%
\theta _{A},\phi _{A},\theta _{B},\phi _{B}$ i.e.

\begin{equation}
p=p(\theta _{A},\phi _{A};\theta _{B},\phi _{B}),\text{ \ \ }q=q(\theta
_{A},\phi _{A};\theta _{B},\phi _{B}).
\end{equation}%
That is, the two-parameter payoff relations can be expressed in the form of (%
\ref{NewGame1}) but this comes at the price that the players' strategic
veriables $p$ and $q$ are not local anymore. For the quantum game with a
one-parameter set of strategies, this is indeed the case as we have $%
p=p(\theta _{A},\phi _{A})$ and $q=q(\theta _{A},\phi _{A}).$

\section*{Conclusion}

We suggest a direct route to obtaining a non-factorizable game from a
classical factorizable game while taking into consideration the quantum
probabilities and the players' strategic variables. We explore how a quantum
game can be considered as a non-factorizable game by considering the scheme
of Eisert et al. for quantization of a bimatrix game that involves four
quantum probabilities. When these probabilities are factorizable in a
specific sense, as described by Eqs.\textit{~}(\ref{factorizability}), the
quantum game attains the interpretation of a mixed-strategy classical game.
This paper presents two approaches in obtaining bimatrix games with
non-factorizable probabilities. Our first approach discusses a
non-factorizable probability distribution in which the non-factorizability
is controlled by an external parameter $k$.

We note that the relations $p=\epsilon _{1}+\epsilon _{2}$ and\ $q=\epsilon
_{1}+\epsilon _{3}$ as obtained from the factorizability constraints (\ref%
{factorizability}) are apparently the simplest expressions connecting the
players' strategies $p$ and $q$ to the quantum probabilities $\epsilon _{i}$ 
$(1\leq i\leq 4)$. However, this does not mean that these are the only
possible expressions that are consistent with the factorizability
constraints (\ref{factorizability}). There can be other possible cases, for
instance, when $p=p(\epsilon _{i})$ and $q=q(\epsilon _{i})$ i.e. both $p$
and $q$ are functions of all four quantum probabilities $\epsilon _{i}$ $%
(1\leq i\leq 4)$. In such a case, the factorizability constraints will take
the following form

\begin{equation}
\epsilon _{1}=p(\epsilon _{i})q(\epsilon _{i}),\text{ }\epsilon
_{2}=p(\epsilon _{i})\left[ 1-q(\epsilon _{i})\right] ,\text{ }\epsilon _{3}=%
\left[ 1-p(\epsilon _{i})\right] q(\epsilon _{i}),\text{ }\epsilon _{4}=%
\left[ 1-p(\epsilon _{i})\right] \left[ 1-q(\epsilon _{i})\right] ,
\label{Factorizability1}
\end{equation}%
and, depending on how the functions $p=p(\epsilon _{i})$ and $q=q(\epsilon
_{i})$ are defined, the analysis will lead to a different outcome of the
quantum game. Essentially, in this paper we use the factorizability
constraints (\ref{factorizability}) to obtain such functions that express
players' strategies in terms of the probabilities $\epsilon _{i}$ and there
can be more than one possible ways of doing that and each way has to respect
the requirement that when $\epsilon _{i}$ are factorizable, now in the sense
of Eqs.~(\ref{Factorizability1}), the quantum game reduces itself to the
classical mixed-strategy game.

This paper addresses several questions relevant to the quantization scheme
proposed by Eisert et al.~in 1999. This scheme motivated other quantization
schemes and has been the basis of a large number of research articles that
have followed since then along this line of research. By systematically
discussing the structure of the payoff relations obtained in this scheme,
and its relation to the corresponding classical mixed-strategy factorizable
game, this paper presents an understanding of the role of quantum
probabilities, factorizability of a probability distribution, and the nature
of the players' strategic variables. We provide a perspective by which a
non-factorizble game is obtained directly from a classical factorizable game
by defining a function $\epsilon _{1}(p,q)$ that satisfyies certain
constraints. The quantum game of Eisert et al. involving one parameter
strategy set is explained as the classical factorizable game. We then study
functions $\epsilon _{1}(p,q)=(pq)^{2}$ and $\epsilon _{1}(p,q)=p^{2}q^{3}$
that satisfy these constraints and lead to non-factorizable games. We
determine the corresponding Nash equlibria. We then ask whether the quantum
game with two-parameter set of strategies can be considered as a
non-factorizable game for a particular choice of the function $\epsilon
_{1}(p,q)$. We find that this can be achieved but it comes at a significant
cost that the players' strategic veriables $p$ and $q$ do not remain local
anymore. This is indeed the case for the quantum game with one-parameter set
of strategies for which the players' strategic variables $p$ and $q$ are
definable in terms of local variables i.e. $p=p(\theta _{A},\phi _{A})$ and $%
q=q(\theta _{A},\phi _{A}).$

\end{document}